\documentclass{PoS}
\usepackage{amsmath,amsthm, amssymb, latexsym}
\usepackage{graphicx} 

\newcommand{\bea}{\begin{eqnarray}}
\newcommand{\eea}{\end{eqnarray}}
\newcommand{\beq}{\begin{equation}}
\newcommand{\eeq}{\end{equation}}

\newcommand{\pdir}{p\kern -5.2pt\raise 0.2ex\hbox {/}}
\newcommand{\vdir}{v\kern -5.75pt\raise 0.15ex\hbox {/}}
\newcommand{\kdir}{k\kern -5.75pt\raise 0.15ex\hbox {/}}
\newcommand{\epsdir}{\epsilon\kern -5.0pt\raise 0.15ex\hbox {/}}
\newcommand{\bvdir}{\bar{v}\kern -5.75pt\raise 0.15ex\hbox {/}}
\newcommand{\Ddir}{D\kern -7.75pt\raise 0.20ex\hbox {/}}
\newcommand{\Adir}{A\kern -7.75pt\raise 0.20ex\hbox {/}}
\newcommand{\ldir}{l\kern -5.0pt\raise 0.2ex\hbox{/}}
\newcommand{\varepsdir}{\varepsilon\kern -5.5pt\raise 0.15ex\hbox{/}}

\def\VEV#1{\langle #1 \rangle}

\newcommand{\pslash}{p \!\!\!/}

\newcommand{\MSB}{\overline{\rm MS}}

\newcommand{\lwrsim}{\raise0.3ex\hbox{$<$\kern-0.75em\raise-1.1ex\hbox{$\sim$}}}

\newcommand{\no}{\nonumber \\}

\def\C2#1#2{({\cal C}_2)_{#1}^{#2}}

\def\eq#1{Eq.~(\ref{#1})}

\newcommand{\re}{\mathbb{R}\text{e}}
\newcommand{\tr}{\text{Tr}}

\title{Egalitarian Improvement to Democracy: Quark renormalization constants
  in $N_f=2$ QCD}

\ShortTitle{Quark renormalization constants}

\author{\speaker{K. Petrov}
        \thanks{Address: Laboratoire de Physique Theorique, Universite Paris-sud, Orsay}\\
       E-mail: \email{Konstantin.Petrov@th.u-psud.fr}}%

\abstract{
We present our results on the non-perturbative evaluation of
the renormalization constant for the quark field, $Z_q$, in        Landau gauge
within RI-MOM
scheme. Using three lattice spacing we are able to isolate lattice artefacts
of various origin, both perturbative and non-perturbative. In particular,
the existence of the dimension-two gluon-condensate is discussed, and
confirmed. \\
\vspace{2cm}\centerline{    \includegraphics[width=50mm]{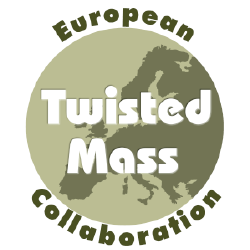}}

}

\FullConference{The XXVIII International Symposium on Lattice Field Theory\\
                July 13-19, 2010\\
                Villasimius, Italy}

\begin{document}

\section{Introduction}
As any other way of computing matrix elements, lattice QCD requires
calculation of the renormalisation constants, which relate bare quantities to
the physical ones. In our particular case the renormalization constants will be
dependent not only on the scale, but also on the lattice parameters, such as light
quark mass and lattice spacing. This dependence turns out to be remarkably
strong in case of the latter.  Even if the lattice
computation contains only $O(a^2)$ lattice artefacts, the bare quantities differ
from the continuum ones by $O(g^2) \simeq O(1/\log(a^2))$ which is unacceptable. 
Renormalization restores the $O(a^2)$ accuracy.  While some perturbative
methods to calculate the constants exist, it is obviously preferable to do it
in a consistent manner, i.e. non-perturbatively. Although in some cases the
results may be similar, there is no guarantee for the validity of the
perturbative calculation. 

Our method of preference is the so-called RI'MOM scheme. It involves the computation of Green
functions  of quarks, gluons, ghosts, at large enough momenta in a fixed gauge,
usually the Landau gauge. This gives the renormalisation constant $Z(\mu)$ at
many values of the scale $\mu$. We have to note here, that even though we
advocate the completely non-perturbative approach, the conversion to other
renormalization schemes has to fall back onto perturbative methods. 
Here we use perturbative QCD to convert MOM into $\MSB$ and run to 2 GeV. The running of
$Z_{\mathrm MOM}(\mu)$ is a very powerful testing tool:  perturbative QCD
is only useful if we are in the perturbative regime, i.e. at large enough
momenta. The only way to check if this is the case is to compare lattice data
with the perturbative running. It turns out that this is not always so.

To calculate the renormalization constants we must fix the gauge, and we
perform all calculation in the Landau gauge. Wilson operator expansion
suggests the presence of the  non-vanishing vacuum expectation value of the
only dimension-two operator in the
Landau  gauge:  $A^2 \equiv A^a_\mu A^{a\mu}$, and that it is not 
small~\cite{Boucaud:2005xn,Blossier:2010ky},\cite{Chetyrkin:2009kh}. This contribution to the OPE will
scale as $1/p^2$  up to logarithmic corrections. To argue that this is a
continuum feature we must check that the $1/p^2$
term  scales well with the lattice spacing when expressed in
physical units. 

In general, our museum of artefacts has a number of exhibitions. The $O(a^2)$ artefacts can be quite large since we consider large
momenta, while finite volume artefacts are minor. There are two main types of
 $O(a^2)$ artefacts: $O(a^2p^2)$ artefacts which respect the continuum
$O(4)$ rotation symmetry, and the ones which do not. The latter appear due to the
explicit breaking of the rotational symmetry, which, on the lattice, is reduced
to the  hypercubic symmetry $H_4$. After removing these, we will identify the $O(a^2p^2)$ 
artefacts non-perturbatively by doing a fit of the running $Z(\mu)$ which will
include the perturbative running, the  condensate
and a term proportional to $a^2p^2$. 

For the elimination of hypercubic artefacts several methods have been proposed in
literature: the democratic one, the perturbative correction and the
non-perturbative ``egalitarian'' one. The perturbative one deserves a separate
paper, so we will focus on two others. We will demonstrate to what extent the ``half-fishbone''
structure, which raw lattice results for $Z_q$ always exhibit and which
is a dramatic illustration of hypercubic artefacts, is corrected by every 
method. 

Although  the issues raised here concern all the renormalisation constants as
well as the QCD coupling constant, we will concentrate in the following on   $Z_q$, that
renormalizes the quark field 
\bea
q_R = Z_q^{1/2} q_B
\eea
where $q_B$ ($q_R$) is the bare (renormalized) quark field. 

\section{Simulation details}

For the details about twisted mass and tree-level improved
Symanzik  gauge actions please refer to refs.~\cite{Boucaud:2007uk,Boucaud:2008xu,Urbach:2007rt,Dimopoulos:2008sy}. 
Here we will just briefly mention the essential parts for our presentation (See Tab.~\ref{setup}).

The Wilson twisted mass fermionic lattice action for two flavours of mass
degenerate  quarks reads  (in the so called twisted
basis~\cite{Frezzotti:2000nk,Frezzotti:2003ni} ) 
\begin{align}
  \label{eq:Sf} 
  \begin{split} 
    S_\mathrm{tm}^{\rm F} = &\, a^4\sum_x\Bigl\{ 
    \bar\chi_x\left[D_{\rm W}+ m_0 + i\gamma_5\tau_3\mu_q  
    \right]\chi_x\Bigr\}\, , \\ 
    & D_{\rm W} = \frac{1}{2}\gamma_\mu\left(\nabla_\mu+\nabla_\mu^{*}\right) 
    -\frac{ar}{2}\nabla_\mu\nabla_\mu^{*} \, ,
  \end{split} 
\end{align}
where $m_0$ is the bare untwisted quark mass and $\mu_q$ the bare twisted  quark
mass, $\tau_3$ is the third Pauli matrix acting in flavour space   and $r$ is
the Wilson parameter, which, as usually, is set to $r=1$ in the simulations. The twisted
 Dirac operator is defined as
\bea\label{eq:DIRtw}
D_{\rm tw} \equiv D_{\rm W}+ m_0  + i\gamma_5\tau_3\mu_q
\eea

The bare quark mass $m_0$
is related as usual to the so-called hopping   parameter $\kappa$, by
$\kappa=1/(8+2am_0)$. Twisted mass fermions are said to be at   {\em maximal
twist} if the bare untwisted mass is tuned to its critical   value,
$m_\mathrm{crit}$. This corresponds to setting the so-called untwisted
PCAC mass to zero, so it requires additional tuning during the simulation.

In the gauge sector  the  tree-level Symanzik improved 
gauge action (tlSym)~\cite{Weisz:1982zw} is used:  
\beq 
  \label{eq:Sg} 
    S_g =  \frac{\beta}{3}\sum_x\Biggl(  b_0\sum_{\substack{ 
      \mu,\nu=1\\1\leq\mu<\nu}}^4\{1-\re\tr(U^{1\times1}_{x,\mu,\nu})\}\Bigr.  
     \Bigl.+ 
    b_1\sum_{\substack{\mu,\nu=1\\\mu\neq\nu}}^4\{1 
    -\re\tr(U^{1\times2}_{x,\mu,\nu})\}\Biggr)\, , 
\eeq
where $\beta \equiv 6 / g_0^2$, $g_0$ being the bare lattice  coupling and we set    $b_1=-1/12$ (with $b_0=1-8b_1$ as dictated by the requirement  of
continuum limit normalization). The overview of the ensembles used for this
work can be
found in Tab.~\ref{setup}.
From our previous simulations we deduced that fortunately $Z_q$ does not depend within
uncertainties on the sea-quarks mass and on the lattice volume. So here we
keep both quark mass and volume fixed.

\begin{table}[ht]
\centering
\begin{tabular}{||c|c|c|c|c||}
\hline
\hline
$\beta$ & $a \mu_q$ & Volume & \# confs dim 3 & \# confs dim 4.
\\ \hline
$3.9$ &  
\begin{tabular}{c}
0.004 \\
0.0064 \\
0.010
\end{tabular}
& 
$24^3\times48$ & 
\begin{tabular}{c}
$120$ \\
$20$ \\
$20$
\end{tabular}
& 
\begin{tabular}{c}
$100$ \\
$20$ \\
$20$
\end{tabular}

\\ \hline
$4.05$ &  
\begin{tabular}{c}
0.006 \\
0.003 \\
0.006 \\
0.008 \\
0.012
\end{tabular}&
\begin{tabular}{c}
 $24^3\times48$  \\
 $32^3\times64$   \\
 $32^3\times64$   \\ 
 $32^3\times64$   \\
 $32^3\times64$ 
\end{tabular}&
\begin{tabular}{c}
$100$ \\
$20$ \\
$20$ \\
$20$ \\
$20$\end{tabular}
& 
\begin{tabular}{c}
$100$ \\
- \\
- \\
-\\
-
\end{tabular}
\\ \hline
$4.2$ & 
\begin{tabular}{c}
0.0065   \\
0.002   \\
\end{tabular}
&\begin{tabular}{c}
 $32^3\times64$ \\
 $24^3\times48$ \\
\end{tabular}
&
\begin{tabular}{c}
$100$\\
-\\
\end{tabular}
&\begin{tabular}{c}
$100$\\
$100$\\
\end{tabular}
\\\hline
\hline
\end{tabular}
\caption{ETMC ensembles used for this work, ``dim 3"
refers to the local operators (V, A, S, P) while ``dim 4" refer to the operators
with one derivative (O44, O14, etc).}
\label{setup}
\end{table}

To compute the renormalization constants for the quark propagator we need to
fix the gauge and calculate the 
 2-point quark 
Green functions on each configuration.
We do this using  a local source taken at a random point on
the lattice which reduces the correlation between successive
configurations:
\bea
S(y,x_0)^{a,\alpha;b_0,\beta_0}_i = D_{\rm tw}^{-1}(y,x)^{a,\alpha;b,\beta} 
so^{b,\beta}(x)\qquad so^{b,\beta}_{i}(x) = \delta_{x,x_0} \delta_{b,b_0}
\delta_{\beta,\beta_0}
\eea
where $i=u,d$ labels isospin.
 Thereafter we perform the Fourier transform of the incoming quark which is a $12\times12$ complex matrix
 \bea\label{Sp}
 S_i(p) \equiv \sum_{y} e^{-i p (y-x_0)} \,S_i(y,x_0)
 \eea
and the outgoing quark  
\bea
S^{\dagger 5}_{i}(p) = \gamma_5 S^{\dagger}_{\bar i}(p) \gamma_5
\eea
where $\bar u\equiv d; \bar d\equiv u$.
We define the {\bf quark renormalization constant $Z_q$} as
\bea\label{Zq}
Z_q(p)\equiv \frac 1 {12\, p^2} \;<{ \rm Tr} [S^{-1}(p) \pslash]> 
\eea
where $<...>$ means here the average over the chosen ensemble of thermalized
configurations.

\section{Artefacts}

\subsection{Hypercubic $H(4)$-extrapolation}
\label{NPhyp}
     
A first kind of artefacts that can be systematically 
cured~\cite{Becirevic:1999uc,deSoto:2007ht} are those arising from the
breaking of the  rotational symmetry of the Euclidean space-time when using an
hypercubic lattice,  where this symmetry is restricted to the discrete $H(4)$
isometry group. It is convenient to compute first the average  of any
dimensionless lattice quantity $Q(a p_\mu)$ over every orbit of the group
$H(4)$. In general several orbits of $H(4)$ correspond to one value of $p^2$.
Defining the $H(4)$ invariants
 \beq
 p^{[4]}=\sum_{\mu=1}^{4} p_\mu^4\qquad p^{[6]}=\sum_{\mu=1}^{6} p_\mu^6
 \eeq
it happens that the orbits of $H(4)$ are 
labeled
by the set $p^2, a^2 p^{[4]},
a^4 p^{[6]}$. 
Group theory tells us that any $H(4)$-invariant 
polynome will only depend on the four 
invariants $p^{[2 i]}$ with $i=1,2,3,4$~\cite{Becirevic:1999uc,deSoto:2007ht}. 
As a consequence of the upper cut for momenta, the first three of these invariants suffice 
to label all the orbits we deal with and hence any presumed dependence on $p^{[8]}$ is  
neglected. We have to admit that our action is obviously non-polynomial, but
perturbative results suggest that up to a certain order the renormalization
constant will be such, so it is sensible to assume this feature. Later we will
be able to check that it leads to consistent results, proving our assumption.
Moreover, in the continuum limit the effect of $ a^2 p^{[4]},
a^4 p^{[6]}$ vanishes. We can thus define the quantity $Q(a p_\mu)$ averaged over 
$H(4)$ as
\beq\label{Q246}
Q(a^2\,p^2, a^4p^{[4]}, a^6 p^{[6]}, a^2\Lambda_{\rm QCD}^2).
\eeq

If the lattice spacing is small enough such that $\epsilon=a^2 p^{[4]}/p^2 \ll 1$,
the dimensionless lattice correlation function defined in~\eq{Q246} can be
expanded in powers of $\epsilon$ and truncated:
\beq
Q(a^2\,{p}^2, a^4p^{[4]}, a^6 p^{[6]},a^2\Lambda_{\rm QCD}^2)
= Q(a^2p^2,a^2\Lambda_{\rm QCD}^2) +
\left.\frac{dQ}{d\epsilon}\right|_{\epsilon=0} a^2
\frac{p^{[4]}}{p^2} + \cdots
\label{eq:p4expan}
\eeq
The basic method is to fit
from the whole set of orbits, sharing the same $p^2$, the
coefficient $dQ/d\epsilon$ and get the extrapolated value of $Q$,
free from $H(4)$ artefacts. 
The fit is performed in a
momentum window between $(p-\delta,p+\delta)$ to extract the extrapolated value
of $Q$ for the momenta $p$ in the window, and then we shift to the next window  etc. 

Such method, which we call SWF, ``sliding window fit'', is quite reliable since the extrapolation does not rely on any  particular
assumption for the coefficients of the Taylor expansion. We can then estimate the systematic
error by varying the width of the
fitting window. 
If we further assume that the coefficient 
\beq
\left.R(a^2p^2,a^2\Lambda_{\rm QCD}^2) = 
\frac{dQ\left(a^2p^2 ,0,0,a^2\Lambda_{\rm
      QCD}^2\right)}{d\epsilon}\right|_{\epsilon=0}
\label{eq:swfexpan}
\eeq
has a smooth dependence on $a^2p^2$ over a given momentum window, we can 
expand $R$ as 
\beq
R=c_{a2p4} +c_{a4p4} a^2p^2 
\eeq
and make a global fit over a wide range of momenta:

\beq
Q(a^2\,{p}^2, a^4p^{[4]}, a^6 p^{[6]},a^2\Lambda_{\rm QCD}^2)
= Q(a^2p^2,a^2\Lambda_{\rm QCD}^2) +
c_{a2p4} a^2 \frac{p^{[4]}}{p^2} + c_{a4p4} a^4p^{[4]}+\cdots
\label{eq:owfexpan}
\eeq
We will refer to this method as OWF, ``one window fit''.

\section{Lattice results and Hypercubic corrections} 
\label{sec:Zqhyp}

\begin{figure}[hbt]
  \hspace{-2cm} 
  \begin{center}
    \begin{tabular}{cc}
      \includegraphics[width=75mm]{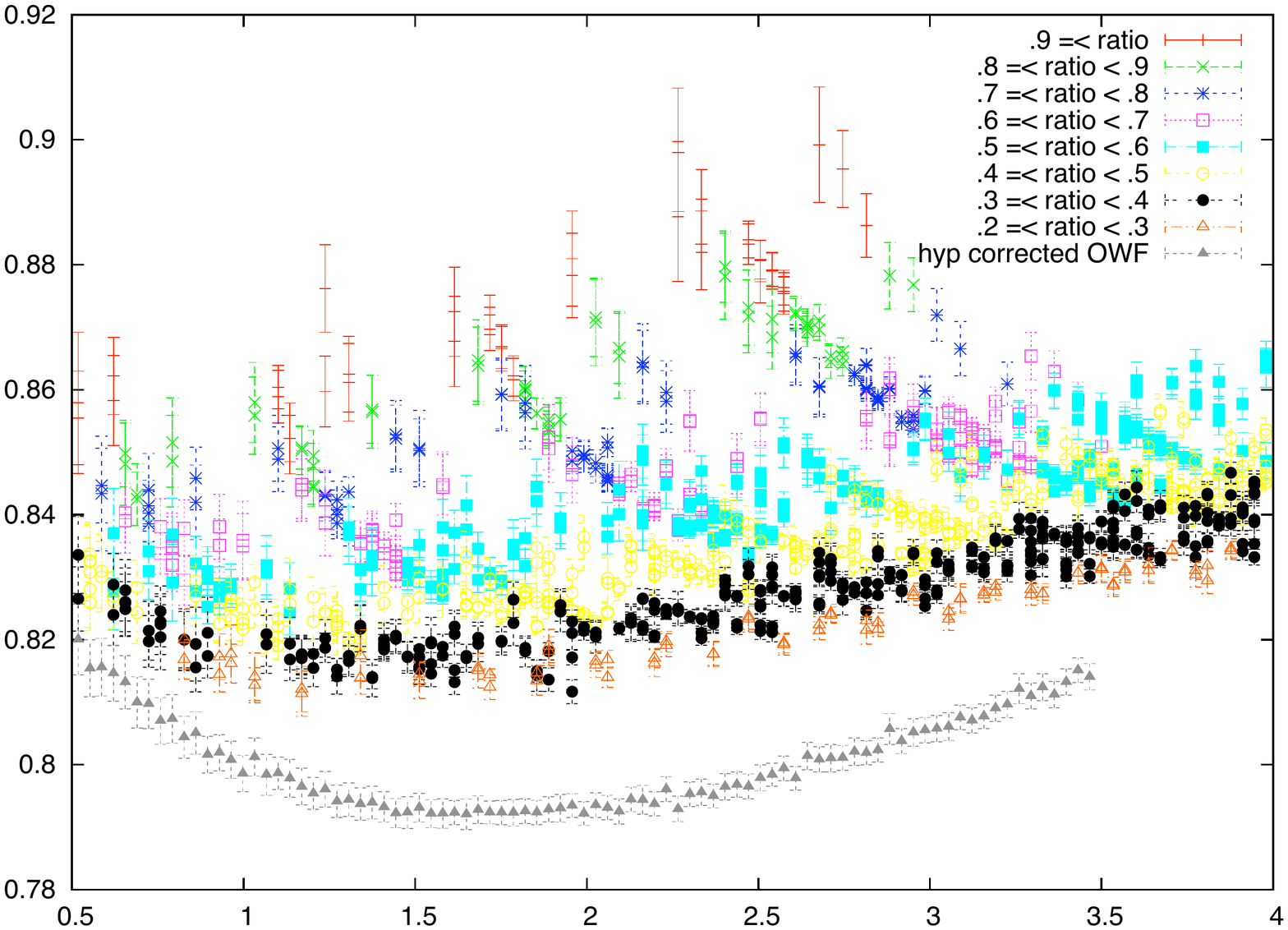}
      
      &
      \includegraphics[width=85mm]{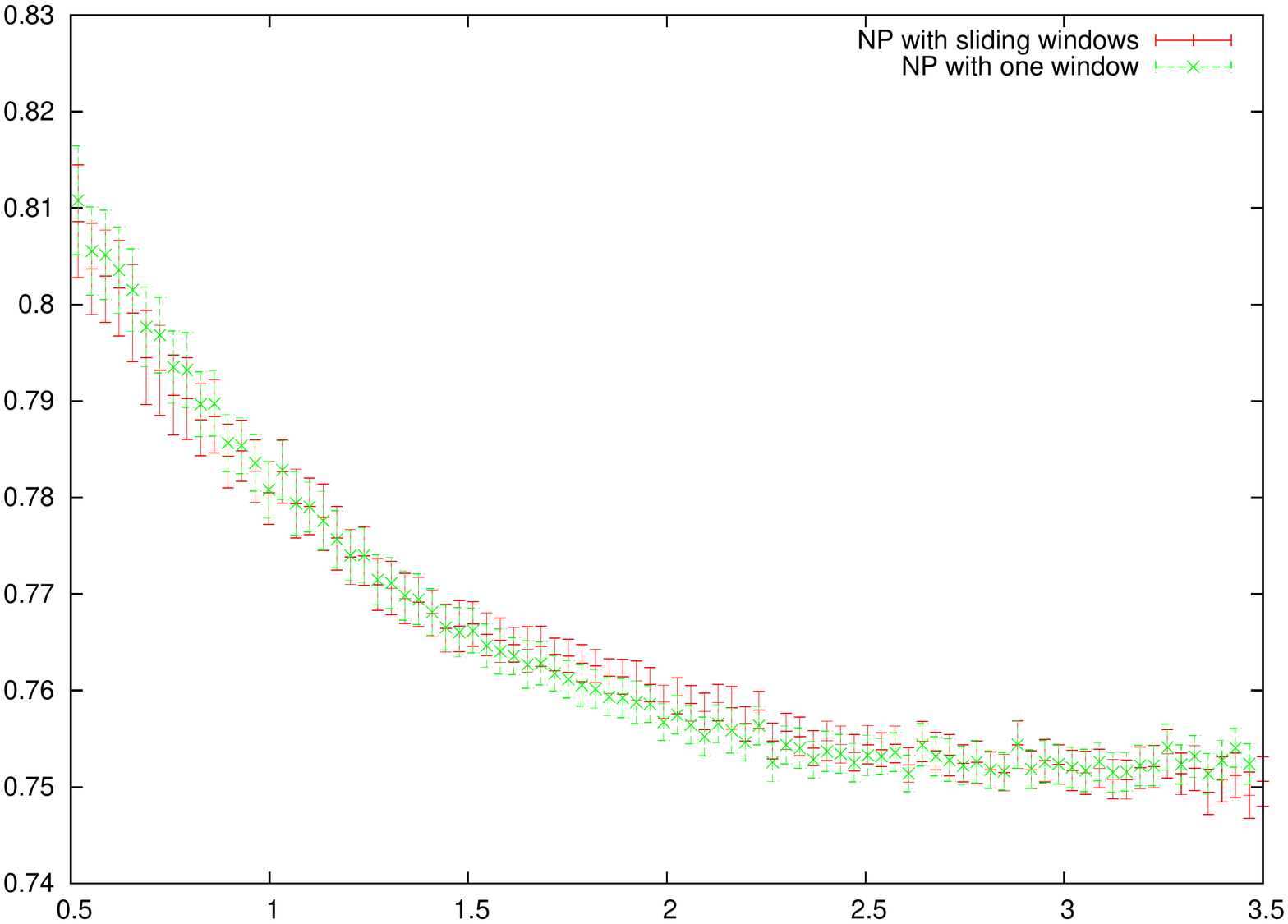}	
      
    \end{tabular}
  \end{center}
  \caption{ 
On the l.h.s. we show the raw data for 
$\beta=3.9$, $Z_q^{\mathrm{latt}}(a^2\,p^2, a^4p^{[4]}, a^6 p^{[6]},  ap_4, a^2\Lambda_{\rm QCD}^2)$. 
 The ``half-fishbone structure'' due to hypercubic artefacts is
     clearly seen. There is one point for every cubic (3-D) orbit.  On
     the r.h.s we show, using the OWF, the non-perturbatively corrected
     data. 
    As on l.h.s., there is one point for every cubic (3-D) orbit.
}
  \label{fig:NP_subtracted}
  \label{fig:fishbone}  
\end{figure}

The hypercubic artefacts are clearly visible on the raw lattice data for  $Z_q^{\mathrm{latt}}$ as the so-called ``half-fishbone''  structure~\cite{Boucaud:2003dx} shown
in fig~\ref{fig:fishbone}.   The color code shows the value of the
ratio $p^{[4]}/(p^2)^2$ which is  between 0.25 and 1. The values which are
closer to one are ``least democratic'' or ``tyrannic'' ones. We see, as
expected,  that the tyrannic points
are more affected by the artefact. We also see that the gap between
$Z_q^{\mathrm {latt}}(a^2\,p^2, a^4p^{[4]}, a^6 p^{[6]},  ap_4,  a^2\Lambda_{\rm
QCD}^2)$ at a given $q^2$ can be as large as 0.07, i.e. about 10\%. Taking a
naive average without a correct treatment  of this artefact  would leave a
systematic upward shift of about 5 \%. 

The oldest method to alleviate this problem is the ``democratic selection''. It amounts to  keeping only, say,
the violet points in   Fig.~\ref{fig:fishbone}, or, in case of low statistics, being less restrictive, the
violet and yellow ones. It works to a certain extent, but obviously we still
have an upward shift compared to H4-method, which is not negligible.

While performing systematic treatment using H4-method, we only
expand up to $p^{[4]}$ since the higher order terms turn out to be
negligible.   In the l.h.s of Fig.~\ref{fig:NP_subtracted},  
we show  the comparison of hypercubic corrected data after applying
OWF and SWF for the case  $\beta=3.9$ . The difference does not appear to be large which is rather
encouraging. The OWF gives a slightly smoother result.

%



\begin{table}[h]
\centering
\begin{tabular}{||c|c|c|c|c|c|c||}
\hline
\hline
$\beta$ & $a^2$ fm$^2$ & $c_{a2p4}$ & $c_{a4p4}$ & $c_{a2p4}/g^2$ &
$c_{a4p4}/g^2$ & $\chi^2$/d.o.f
\\ \hline
$3.9$ & 0.00689 & 0.067(4) &-0.0149(10)&0.044(3)&-0.0097(7)& 4.1 \\ \hline
$4.05$ &0.00456& 0.065(3)& -0.0144(5) & 0.044(2)&-0.0097(3)&0.53\\ \hline
$4.2$ &0.00303&  0.055(11)& -0.0124(4)&0.039(8) &-0.0089(3)&0.98 \\ \hline
\end{tabular}
\caption{Results for the slope in $a^2 \,p^{[4]}/p^2$  and $a^4 \,p^{[4]}$ 
and the same divided by $g^2$ in the one window fits.}
\label{tab:p4slopes}
\end{table}



The fitted values of $c_{a2p4}$ and $c_{a4p4}$ from the one window fit are given in 
Tab.~\ref{tab:p4slopes} as well as the same divided by $g^2$, since
perturbation theory expects at least for $c_{a2p4}$ to be $\propto g^2$.
Before dividing by $g^2$ a small scaling violation is apparent, and much less
so afterwards.
 The $\chi^2$ in 
Tab.~\ref{tab:p4slopes} is not good for $\beta=3.9$, apparently due to some
structure at the lower end of the plot,  but remember it uses
only two hypercubic parameters.  Of course, this point is further away
from the continuum than both other points.

\section{Results and Conclusion}

Once the hypercubic corrections are performed we make a fit for every $\beta$ 
according to the following formula
\bea\label{eq:zqfit}
 Z_q^{\mathrm{NPsubtracted}}(a^2p^2)&=&Z_0\,c_{0Z_q}(q^2,\mu^2)\left( 1
+ \frac{c_{1overp2}} {p^2} \right) + 
c_{a2p2}\; a^2\,p^2
\eea
where $c_{0Z_q}(q^2,\mu^2)$ is the perturbative running of 
$Z_q$~\cite{Chetyrkin:1999pq}, 
$c_{a2p2}$ is a hypercubic insensitive lattice artefact, and 
$c_{1overp2}$ is due to the $g^2 \VEV{A^2}$ condensate.
Combining several analysis methods we get 
\bea\label{eq:final}
g^2(\mu^2) \VEV{A^2}_{\mu^2} &=&  2.25(40)\left(^{+0.75}_{-1.0}\right)
 \;\mathrm {GeV}^2  \quad \mu=10\,
\mathrm{GeV}\no
Z_q^{\mathrm {pert}}((10\,{\mathrm {GeV}})^2,g^2_{\mathrm {bare}}) 
&=& 0.741(3)(7) - 0.305(25) \,(g^2_{\mathrm {bare}}-1.5)\no
Z_q^{\mathrm {pert}}((2\,{\mathrm {GeV}})^2,g^2_{\mathrm {bare}}) 
&=& 0.773(3)(7) - 0.319(20) \,(g^2_{\mathrm {bare}}-1.5)\no
\eea


To conclude, we calculated $Z_q$ from the ETMC gauge configurations in the RI-MOM
scheme. We demonstrated effectiveness of the H4-method of removing hypercubic
artefacts. 
From the resulting hypercubic corrected function  
$Z_q(a^2p^2,a^2\Lambda_{\rm QCD}^2)$ we perform a fit to an Ansatz consisting
of the perturbative running, a non perturbative $1/p^2$ term  and
rotationally-symmetric lattice spacing artefact 
proportional to $a^2p^2$. The fits are good and $a^2p^2$ term scales almost
perfectly in lattice units, as expected. The $g^2 \VEV{A^2}$ term scales rather well in
physical units as expected, though less accurate. Therefore we advocate the
presence of the non-perturbative condensate in the quark renormalization
constant. More detailed analysis and even more rigorous proof will be
presented in a separate paper soon.

\end{document}